\documentclass[aps,prl,twocolumn,superscriptaddress,showpacs,preprintnumbers,amsmath,amssymb]{revtex4}

\usepackage{graphicx} 
\usepackage{dcolumn}  
\usepackage{subfigure}

\newcommand{\cc}{\ensuremath{(c\bar{c})_{\mathrm{res}}}}

\newcommand{\RM}{\ensuremath{M_{\mathrm{recoil}}}}
\newcommand{\gevc}{\ensuremath{\, {\mathrm{GeV}/c^2}}}
\newcommand{\mevc}{\ensuremath{\, {\mathrm{MeV}/c^2}}}

\newcommand{\Xmass}{\ensuremath{(3.943 \pm 0.006 \pm 0.006) \, {\mathrm{GeV}}/c^2 }}
\newcommand{\Xwidth}{\ensuremath{52 \, {\mathrm{MeV}}/c^2}}
\newcommand{\Xddst}{\ensuremath{(96^{+45}_{-32} \pm 22) \%}}
\newcommand{\Xdd}{\ensuremath{41  \%}}
\newcommand{\Xom}{\ensuremath{26  \%}}
\newcommand{\Xborn}{\ensuremath{(10.6 \pm 2.5 \pm 2.4)\, \mathrm{fb}}}

\begin{document} 


\title{ \quad\\[0.5cm] \Large Observation of a new charmonium-like
state produced in association with a $J/\psi$ in $e^{+}e^{-}$
annihilation at $\sqrt{s} \approx 10.6~$GeV}

\affiliation{Budker Institute of Nuclear Physics, Novosibirsk}
\affiliation{Chonnam National University, Kwangju}
\affiliation{University of Cincinnati, Cincinnati, Ohio 45221}
\affiliation{University of Frankfurt, Frankfurt}
\affiliation{Gyeongsang National University, Chinju}
\affiliation{University of Hawaii, Honolulu, Hawaii 96822}
\affiliation{High Energy Accelerator Research Organization (KEK), Tsukuba}
\affiliation{Hiroshima Institute of Technology, Hiroshima}
\affiliation{Institute of High Energy Physics, Chinese Academy of Sciences, Beijing}
\affiliation{Institute of High Energy Physics, Vienna}
\affiliation{Institute for Theoretical and Experimental Physics, Moscow}
\affiliation{J. Stefan Institute, Ljubljana}
\affiliation{Kanagawa University, Yokohama}
\affiliation{Korea University, Seoul}
\affiliation{Kyungpook National University, Taegu}
\affiliation{Swiss Federal Institute of Technology of Lausanne, EPFL, Lausanne}
\affiliation{University of Ljubljana, Ljubljana}
\affiliation{University of Maribor, Maribor}
\affiliation{University of Melbourne, Victoria}
\affiliation{Nagoya University, Nagoya}
\affiliation{Nara Women's University, Nara}
\affiliation{National Central University, Chung-li}
\affiliation{National United University, Miao Li}
\affiliation{Department of Physics, National Taiwan University, Taipei}
\affiliation{H. Niewodniczanski Institute of Nuclear Physics, Krakow}
\affiliation{Nippon Dental University, Niigata}
\affiliation{Niigata University, Niigata}
\affiliation{Osaka City University, Osaka}
\affiliation{Peking University, Beijing}
\affiliation{Princeton University, Princeton, New Jersey 08544}
\affiliation{RIKEN BNL Research Center, Upton, New York 11973}
\affiliation{University of Science and Technology of China, Hefei}
\affiliation{Shinshu University, Nagano}
\affiliation{Sungkyunkwan University, Suwon}
\affiliation{University of Sydney, Sydney NSW}
\affiliation{Tata Institute of Fundamental Research, Bombay}
\affiliation{Toho University, Funabashi}
\affiliation{Tohoku Gakuin University, Tagajo}
\affiliation{Tohoku University, Sendai}
\affiliation{Department of Physics, University of Tokyo, Tokyo}
\affiliation{Tokyo Institute of Technology, Tokyo}
\affiliation{Tokyo Metropolitan University, Tokyo}
\affiliation{Tokyo University of Agriculture and Technology, Tokyo}
\affiliation{University of Tsukuba, Tsukuba}
\affiliation{Virginia Polytechnic Institute and State University, Blacksburg, Virginia 24061}
   \author{K.~Abe}\affiliation{High Energy Accelerator Research Organization (KEK), Tsukuba} 
   \author{I.~Adachi}\affiliation{High Energy Accelerator Research Organization (KEK), Tsukuba} 
   \author{H.~Aihara}\affiliation{Department of Physics, University of Tokyo, Tokyo} 
 \author{K.~Arinstein}\affiliation{Budker Institute of Nuclear Physics, Novosibirsk} 
   \author{Y.~Asano}\affiliation{University of Tsukuba, Tsukuba} 
 \author{V.~Aulchenko}\affiliation{Budker Institute of Nuclear Physics, Novosibirsk} 
   \author{T.~Aushev}\affiliation{Institute for Theoretical and Experimental Physics, Moscow} 
   \author{T.~Aziz}\affiliation{Tata Institute of Fundamental Research, Bombay} 
   \author{A.~M.~Bakich}\affiliation{University of Sydney, Sydney NSW} 
 \author{V.~Balagura}\affiliation{Institute for Theoretical and Experimental Physics, Moscow} 
% ...
\collaboration{The Belle Collaboration}

\noaffiliation

\begin{abstract}
We report the first observation of a new charmonium-like state at a mass of
\Xmass . This state, which we denote as $X(3940)$, is observed in the
spectrum of masses recoiling from the $J/\psi$ in the inclusive
process $e^+ e^- \to J/\psi + \rm{anything}$. We also observe its
decay into $D^* \overline{D}$ and determine its intrinsic width to be
less \Xwidth\ at the 90\% C.L. These results are obtained from a
$357\,{\rm fb}^{-1}$ data sample collected with the Belle detector
near the $\Upsilon(4S)$ resonance, at the KEKB asymmetric-energy $e^+
e^-$ collider.
\end{abstract}

\pacs{13.66.Bc,12.38.Bx,14.40.Gx}

\maketitle
\setcounter{footnote}{0}

Recently there have been a number of reports of new charmonium or
charmonium-like states: $\eta_c(2S)$~\cite{choi1},
$X(3872)$~\cite{choi2}, $Y(3940)$~\cite{choi3} and
$Y(4260)$~\cite{baby}. The latter three states have not been assigned
to any charmonium states in the conventional quark model. Moreover,
charmonium production in different processes is not well
understood. One striking example is the surprisingly large cross
section for double charmonium production in $e^+ e^-$ annihilation
observed by Belle~\cite{2cc} and confirmed by BaBar~\cite{bab}.  These
experimental results have generated renewed theoretical interest in
the spectroscopy, decays and production of charmonium.

In this paper we report the observation of a new charmonium-like state
above $D \overline{D}$ threshold, $X(3940)$, produced in the process
$e^+ e^- \to J/\psi\, X(3940)$.  We also present results from searches
for $X(3940)$ decay into $D\overline{D}$, $D^* \overline{D}$ and
$J/\psi \omega$. The data used for this analysis correspond to an
integrated luminosity of $357\,\mathrm{ fb}^{-1}$ collected by the
Belle detector at the $\Upsilon(4S)$ resonance and nearby continuum at
the KEKB asymmetric-energy $e^+ e^-$ collider.


The $J/\psi$ reconstruction procedure is identical to our previously
published analyses~\cite{2cc, 2cc2}. Oppositely charged tracks that
are positively identified as muons or electrons are used for $J/\psi
\to \ell^+ \ell^-$ reconstruction. A partial correction for final
state radiation and bremsstrahlung energy loss is performed by
including the four-momentum of every photon detected within a
$50\,\mathrm{mrad}$ cone around the electron direction in the
$e^{+}e^{-}$ invariant mass calculation. The two lepton candidate
tracks are required to have a common vertex, with a distance from the
interaction point in the plane perpendicular to the beam axis smaller
than $1\,\mathrm{mm}$. The $J/\psi \to \ell^{+}\ell^{-}$ signal region
is defined by the mass window $\left|M_{\ell^{+} \ell^{-}} -
M_{J/\psi}\right| < 30 \, \mathrm{ MeV}/c^2$ ($\approx 2.5\, \sigma$).
$J/\psi$ candidates in the signal window are subjected to a mass and
vertex constrained fit to improve their momentum resolution.  QED
processes are substantially suppressed by requiring the total charged
multiplicity ($N_{\rm ch}$) in the event to be greater than four.
Background due to $J/\psi$ mesons produced from $B\overline{B}$ events
is removed by requiring the center-of-mass (CM) momentum
$p^{*}_{J/\psi}$ to be greater than $2.0\,\mathrm{GeV}/c$. As in the
previous analysis, we define the recoil mass as
\begin{equation}
M_{\rm recoil}(J/\psi) = \sqrt{(E_{\rm
CMS}-E_{J/\psi}^*)^2-p_{J/\psi}^{*~2}},
\end{equation}
where $E^*_{J/\psi}$ is the $J/\psi$ CM energy after the mass
constraint.

For the study of the $X(3940)\to D^{(*)}\overline{D}$, we reconstruct
$D^0$ candidates using five decay modes: $K^- \pi^+$, $K^- K^+$, $K^-
\pi^- \pi^+ \pi^+$, $K_S^0 \pi^+ \pi^-$ and $K^- \pi^+ \pi^0$; and
$D^+$ candidates using $K^- \pi^+ \pi^+$, $K^- K^+ \pi^+$ and $K_S^0
\pi^+$.  For the $D^0 \to K^- \pi^- \pi^+ \pi^+$ and $D^0 \to K^-
\pi^+ \pi^0$ modes, mass windows of $\pm 10\mevc$ and $\pm 20\mevc$
are used; a $\pm 15\mevc$ mass window is used for all other modes
(approximately $2.5\,\sigma$ in each case). To improve their momentum
resolution, $D$ candidates are refitted to the nominal $D^0$ or $D^+$
masses. To study the contribution of combinatorial background under
the $D$ peak, we use $D$ sidebands with mass windows that are four
times as large. For the $X(3940)\to J/\psi \, \omega$ search,
candidate $\omega$ mesons are reconstructed from $\pi^+ \pi^- \pi^0$
combinations within $\pm 20\mevc$ ($\sim 2.5\,\sigma$) of the nominal
$\omega$ mass. The $\omega$ sideband region is defined by $30< |
M(\pi^+ \pi^- \pi^0) - M_{\omega}| < 50\mevc$.

\begin{figure}[htb]
\hspace*{-0.025\textwidth}
\includegraphics[width=0.48\textwidth] {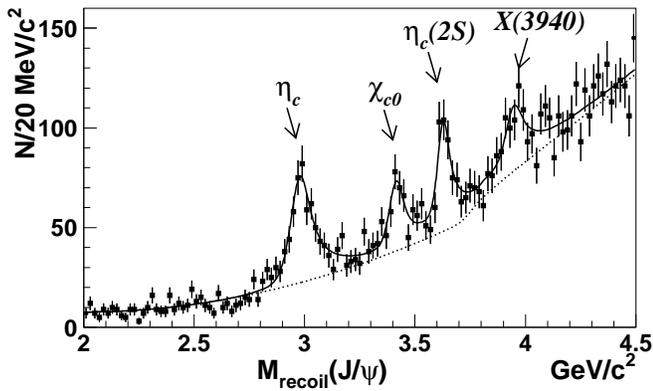}
\caption{The distribution of masses recoiling against the
reconstructed $J/\psi$ in inclusive $e^+e^- \to J/\psi X$ events.  
The curves are described in the text.}
\label{cc4}
\end{figure}

The recoil mass spectrum for the inclusive $J/\psi$ event sample is
shown in Fig.~\ref{cc4}.  Here, in addition to the three previously
reported peaks at the $\eta_c$, $\chi_{c0}$ and $\eta_c(2S)$ masses,
there is a fourth peak above $D\overline{D}$ threshold.
We perform a fit to this spectrum that includes the three previously
seen charmonium states plus a fourth state with mass near
$3.94\gevc$. The expected signal line-shapes are determined from Monte
Carlo (MC) simulation as described in previous Belle
publications~\cite{2cc,2cc2}. The mass values for the $\eta_c$,
$\chi_{c0}$, $\eta_c(2S)$ and $X(3940)$ are free parameters in the
fit, the widths of $\eta_c$ and $\chi_{c0}$ are fixed to PDG
values~\cite{pdg} and the $\eta_c(2S)$ width is fixed to
$\Gamma=17\mevc$~\cite{bab_eta2}. The width of the new state is a free
parameter in the fit. The signal function for the $X(3940)$ is a
convolution of the (zero-width) MC line shape with a Breit-Wigner
function.  The background is parameterized by a second order polynomial
and a threshold term ($\sqrt{\RM(J/\psi)-2M_D}$) to account for a
possible contribution from $e^+ e^- \to J/\psi D^{(*)}
\overline{D}{}^{(*)}$.

The fit results are given in Table~\ref{cc3t}.  The significance for
each signal is defined as $\sqrt{-2\ln
(\mathcal{L}_0/\mathcal{L}_{\text {max}})}$, where $\mathcal{L}_0$ and
$\mathcal{L}_{\text {max}}$ denote the likelihoods returned by the
fits with the signal yield fixed at zero and at the fitted value,
respectively. The significance of the $X(3940)$ signal is
$5.0\,\sigma$.  The fitted width of the $X(3940)$ state is consistent
with zero within its large statistical error: $\Gamma= 39 \pm 26
\mevc$.  The fit results are shown in Fig.~\ref{cc4} as the solid
curve; the dashed curve is the background function.
\begin{table}[htb]
\caption{Summary of the signal yields, charmonium masses and
significances for $e^+ e^- \to J/\psi \, \cc$.}
\label{cc3t}
\begin{center}
\begin{tabular}
{@{\hspace{0.4cm}}l@{\hspace{0.4cm}}||@{\hspace{0.4cm}}c@{\hspace{0.4cm}}|@{\hspace{0.4cm}}c@{\hspace{0.4cm}}|@{\hspace{0.4cm}}c@{\hspace{0.4cm}}}
\hline \hline 
\cc & $N$ & $M\,[\mathrm{GeV}/c^2]$& $N_\sigma$ \\ \hline \hline
$\, \eta_c$ & $\, 501 \pm 44 \,$ & $\, 2.970 \pm 0.005 \,$ & $\, 15.3 \,$ \\

$\, \chi_{c0}$ & $230 \pm 40 $ & $3.406 \pm 0.007 $ & 6.3  \\

$\, \eta_c(2S)$ & $311 \pm 42$ & $3.626 \pm 0.005 $ & 8.1 \\

$\, X(3940)$ & $266 \pm 63$ & $3.936 \pm 0.014 $ &  5.0 \\ 
\hline \hline
\end{tabular}
\end{center}
\end{table}

\begin{figure}[htb]
\includegraphics[width=0.48\textwidth]{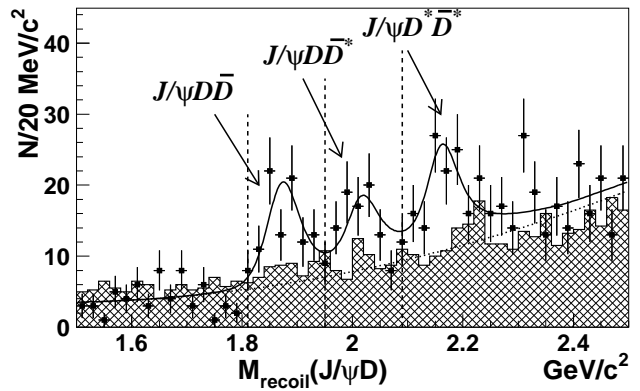}
\caption{The distribution of masses recoiling against the
reconstructed $J/\psi \,D$ combinations in the data.  Points with
error bars correspond to the $D$ mass signal window; hatched
histograms show the scaled $D$ sideband contribution. The solid line
represents the fit described in the text. The dashed line shows the
background function. }
\label{psiD} 
\end{figure}

The new state has a mass that is above both the $D\overline{D}$ and
$D^*\overline{D}$ thresholds.  We therefore perform a search for $X(3940)$
decays into $ D\overline{D}$ and $ D^*\overline{D}$ final
states. Because of the small product of $D^{(*)}$ reconstruction
efficiencies and branching fractions, it is not feasible to
reconstruct fully the chain $e^+ e^- \to J/\psi \, X(3940), \, \,
X(3940) \to D^{(*)}\overline{D}$. To increase the efficiency, only one
$D$ meson in the event is reconstructed in addition to the
reconstructed $J/\psi$ and the other $\overline{D}$ or
$\overline{D}{}^{*}$ is detected as a peak in the spectrum of masses
recoiling against the $J/\psi D$ combination. The Monte Carlo
simulation for $e^+ e^- \to J/\psi D \overline{D}$ and $e^+ e^- \to
J/\psi D^* \overline{D}$ processes indicates a \RM$(J/\psi D)$
resolution of about $30\mevc$ and a separation between these two
processes of $2.5\, \sigma$. Figure~\ref{psiD} shows the \RM$(J/\psi
D)$ spectrum in the $D$ mass window (points with error bars) and in
the scaled $D$ mass sidebands (hatched histogram), where $D$ includes
$D^0$ and $D^+$. Some events have multiple $D$ candidates. In these
cases, only the candidate with invariant mass closest to the nominal
$D$-meson mass is used. Two peaks around the nominal $D$ and $D^*$
masses are clearly visible in this distribution. The excess of real
$D$ events compared to the $D$ sidebands at masses above $2.1\gevc$ is
due to $e^+ e^- \to J/\psi D^* \overline{D}{}^*$ or $e^+ e^- \to
J/\psi D^{(*)} \overline{D}{}^{(*)} \pi$ processes. A fit to this
spectrum is performed using shapes fixed from MC for three processes
($J/\psi D \overline{D}$, $J/\psi D^* \overline{D}$ and $J/\psi D^*
\overline{D}{}^*$) and a second order polynomial.
The fit gives $N_{D\overline{D}}=86 \pm 17$ ($5.1 \, \sigma $) and
$N_{D^*\overline{D}}=55 \pm 18$ ($3.3 \, \sigma$) events in the $D$
and the $D^*$ peaks, respectively.  Selecting events from the
$\RM(J/\psi D)$ regions around the $D$ and $D^*$ masses ($\pm
70\mevc$), we thus effectively tag the processes $e^+ e^- \to J/\psi D
\overline{D}$ and $e^+ e^- \to J/\psi D^* \overline{D}$.

We constrain $\RM(J/\psi D)$ to the $D^{(*)}$ nominal mass, thereby
improving the $M(D^{(*)}\overline{D})\equiv \RM(J/\psi)$ resolution by
a factor of 2.5 ($\sigma \sim 10\mevc$ after constraint), according to
the MC simulation. In the $X(3940) \to D^* \overline{D}$ case, the
reconstructed $D$ can be from either the $X(3940)$ decays or the $D^*$
decay: the constraint $\RM(J/\psi D) \to M(D^*)$ also works in the
latter case, as both $X(3940)\to D^* \overline{D}$ and $D^*$ decays
have very little available phase space.

\begin{figure}[htb]
\includegraphics[width=0.48\textwidth]{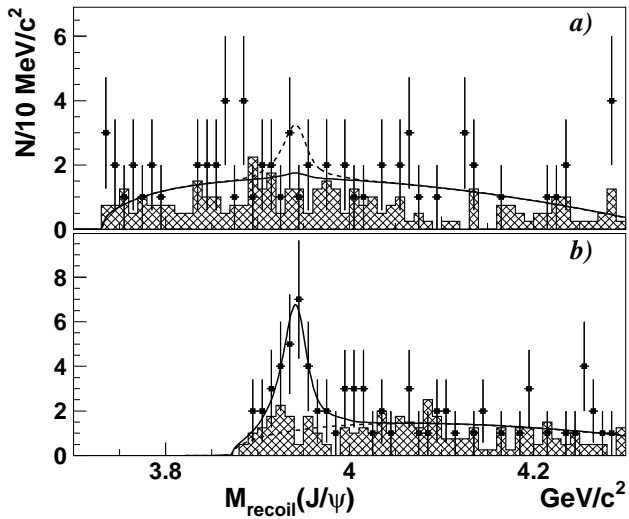}
\caption{The $\RM(J/\psi)$ distribution for events tagged and
constrained as a) $e^+ e^- \to J/\psi D \overline{D}$, and b) $e^+ e^-
\to J/\psi D^* \overline{D}$.  The hatched histograms correspond to
scaled $D$ sidebands. The solid lines are result of the fits,
described in the text. The dashed lines show: a) the 90\% C.L.  upper
limit on the signal; b) the background function.}
\label{psi_rm_d} 
\end{figure}

The resulting $\RM(J/\psi)$ distributions are shown in
Figs.~\ref{psi_rm_d}\,a) ($M(D)$ region) and ~\ref{psi_rm_d}\,b)
($M(D^*)$ region). The cross-hatched histograms show the scaled $D$
sideband distributions. For events with multiple entries, the
candidate with invariant mass closest to the nominal $D$-meson mass
(for $D$ signal window) or closest to the center of the sideband (for
sidebands) is used.
An $X(3940)$ peak with a resolution that is better than that for the
unconstrained $\RM(J/\psi)$ distribution is evident in
Fig.~\ref{psi_rm_d}\,b), corresponding to the decay $X(3940)\to D^*
\overline{D}$. We perform a fit to this distribution.  The signal
function is a convolution of a Breit-Wigner with a free width and a
resolution function fixed to the MC expectation. The background
function is a threshold function $(A+B \cdot
M(D^*\overline{D}))\sqrt{M(D^*\overline{D})-M_{\mathrm{thr}}}$ with
$M_{\mathrm{thr}}\equiv M(D^*)+ M(D)$. The fit yields the number of
signal events in the peak $N=24.5 \pm 6.9$ with a statistical
significance of $5.0 \, \sigma$. The width of the $X(3940)$ is
$\Gamma=(15.4\pm 10.1)\mevc$. The mass of the state is measured to be
$M=(3.943\pm 0.006)\gevc$.

We perform a similar fit to the $\RM(J/\psi)$ distribution for events
tagged and constrained as $e^+ e^- \to J/\psi D \overline{D}$. Since
no $X(3940)$ signal is seen for this mode, we fit this distribution
with $X(3940)$ parameters fixed to the values found by the fit of
tagged $e^+ e^- \to J/\psi D^* \overline{D}$.  The signal yield is
found to be $0.2^{+4.4}_{-3.5}$ events and we set an upper limit for
the $X(3940)$ signal of 8.1 events at the 90\% C.L.

\begin{figure}[htb]
\includegraphics[width=0.48\textwidth]{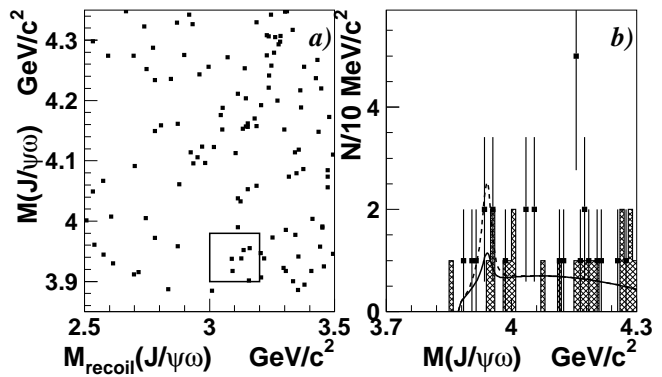}
\caption{a) The scatterplot of $\RM(J/\psi\, \omega)$ {\emph {vs.}}
$M(J/\psi\,\omega)$, and projection onto b) $M(J/\psi \, \omega)$.
Points with errors bars show the contribution from the $\omega$ mass
window; the hatched histogram shows the $\omega$ sideband. The solid
line represents the fit described in the text and the dashed line the
90\% C.L.  upper limit on the $X(3940) \to J/\psi \omega$
contribution.}
\label{omega} 
\end{figure}

An enhancement with a similar mass, $Y(3940)$, decaying into $J/\psi\,
\omega$ has been recently observed by Belle~\cite{choi3} in $B$
decays. We perform a search for the decay $X(3940) \to J/\psi
\,\omega$ to see if $X(3940)$ and $Y(3940)$ could be the same
particle. To increase the efficiency we reconstruct the $\omega$ and
only one $J/\psi$ from the $J/\psi\, J/\psi\, \omega$ final state.
The unreconstructed $J/\psi$ is identified as a peak in the spectrum
of recoil masses against the reconstructed $J/\psi \,\omega$
combinations.  A signal for $X(3940) \to J/\psi \, \omega$ would be
seen as a peak around the nominal $X(3940)$ mass in a distribution of
$\RM (J/\psi)- \RM(J/\psi \omega)+ M(J/\psi)$ if the reconstructed
$J/\psi$ is prompt, and in the $J/\psi \,\omega$ invariant mass
distribution if the reconstructed $J/\psi$ is from the $X(3940)$
decay. Since the first case has much larger combinatorial background
and less sensitivity, we use only the second case. A scatterplot of
\RM($J/\psi \omega$) {\emph {vs.}}  $J/\psi \,\omega$ invariant mass
in the data is shown in Fig.~\ref{omega}\,a). An $M(J/\psi \,\omega)$
projection with the additional requirement $|\RM(J/\psi \, \omega) -
M_{J/\psi} | < 100 \mevc$ is shown in Fig.~\ref{omega}\,b). A fit to
this distribution is done with the signal function and parameters
fixed from the result of the $D^* \overline{D}$ tagged fit; the
background is a threshold function.  The fit yields
$1.9^{+3.2}_{-2.4}$ signal events corresponding to a $7.4$ event upper
limit at the 90\% C.L.


The systematic errors for the $e^+ e^- \to J/\psi X(3940)$ Born cross
section and for the $X(3940)$ branching fractions
$\mathcal{B}(X(3940))$ are summarized in Table~\ref{sys}.
\begin{table}[]
\caption{Contribution to the systematic error for
$\sigma_{\text{Born}}(e^+ e^- \to J/\psi\, X(3940))$ and $\mathcal{B}
(X(3940))$ [\%]. }
\label{sys}
\begin{center}
\begin{tabular}
{@{\hspace{0.1cm}}l@{\hspace{0.1cm}}||@{\hspace{0.1cm}}c@{\hspace{0.1cm}}|@{\hspace{0.1cm}}c@{\hspace{0.1cm}}|@{\hspace{0.1cm}}c@{\hspace{0.1cm}}|@{\hspace{0.1cm}}c@{\hspace{0.1cm}}}
\hline  \hline
Source & $\sigma_{\text{Born}}$ & 
\multicolumn{3}{c}{$\mathcal{B}(X(3940))$} \\ \cline{3-5}
 & & $D^{*} \overline{D}$  & $D \overline{D}$ & $J/\psi \omega$ \\
\hline
Fitting procedure     & $\pm 11$ & $\pm 17$ & $ - $    & $ - $ \\
Angular distributions & $\pm 19$ & $\pm 12$ & $\pm 12$ & $ \pm 16$  \\
$N_{\mathrm {ch}}$ requirement          & $\pm 3$  & $\pm 3$  & $\pm 3 $ & $ \pm 3 $ \\        
Reconstruction        & $\pm 2$  & $\pm 6$ & $\pm 6$   & $\pm 5$ \\
Identification        & $\pm 3$  & $\pm 1$ & $\pm 1$ & $ - $ \\
\hline
Total                 & ~~$\pm 23$~~ & ~~$\pm 22$~~ & ~~$\pm 14$~~ & ~~$\pm 17$~~ \\
\hline \hline
\end{tabular}
\end{center}
\end{table}
To estimate the systematic errors associated with the fitting
procedure we study the difference in $X(3940)$ yield returned by the
fit to the \RM$(J/\psi)$ distribution under different assumptions for
the signal and background parameterization. In particular, in the
first fit we use a background function that includes several threshold
functions corresponding to the production of $D^*\overline{D}$ and
$D^*\overline{D}{}^*$. We also use the threshold function
$(A+B\cdot\RM(J/\psi))\sqrt{\RM(J/\psi)-M_{\mathrm{thr}}}$.  Different
angular distributions result in different $J/\psi$ (and $D$)
reconstruction efficiencies.  In the MC the $J/\psi$ production angle
and $J/\psi, \, X(3940)$ helicity angle distributions are assumed to
be flat. The possible extreme angular distributions ($1+
\cos^2{\theta}$ and $\sin^2\theta$) are considered to estimate the
systematic uncertainty of this assumption. This uncertainty partially
cancels out in the determination of $\mathcal{B}(X(3940))$ because of
the common $J/\psi$ efficiency.  Other contributions come from
$N_{\mathrm{ch}}>4$ requirement efficiency, track reconstruction
efficiency, lepton identification for reconstructed $J/\psi$ and kaon
identification for reconstructed $D$.

The systematic errors in the measurement of the $X(3940)$ mass are
dominated by the $5\,\mevc$ uncertainty associated with the fitting
procedure. The uncertainty due to the $J/\psi$ momentum scale is less
than $3\,\mevc$~\cite{2cc2}.  These contributions are added in
quadrature to give $6\,\mevc$ total. The systematic error on the mass
of the nearby $\eta_c(2S)$, which is still poorly known~\cite{pdg}, is
found to be the same.  From the fit to Fig.~\ref{psi_rm_d}\,b) we
estimate the $X(3940)$ width to be smaller than $47\,\mevc$ at the
90\% C.L.; this takes into account the fact that the likelihood
function is not parabolic. When fitting systematics are taken into
account, we find $\Gamma < 52\,\mevc$ at the 90\% C.L.


The Born cross section for $e^+ e^- \to J/\psi X(3940)$ is calculated
following the procedure used in Ref.~\cite{2cc2}.
As in the previous Belle papers, because of selection
criteria the result is presented in terms of the product of the cross
section and the branching fraction of the $X(3940)$ into more than two
charged tracks ($\mathcal{B}_{>2}$). We obtain
\begin{eqnarray}
\sigma_{\text {Born}} \times \mathcal{B}_{>2} = \Xborn .
\end{eqnarray}

Using the $X(3940)$ yields in inclusive and $D^{*}\overline{D}$ tagged
samples, we calculate the fraction of $X(3940)$ decays with more than
two charged tracks in the final state into $D^{*}\overline{D}$, 
$\mathcal{B}_{>2}(X(3940) \to D^{*} \overline{D})$.  To remove the
correlation between the two samples, we apply a veto on
$D^{*}\overline{D}$ tagging in the inclusive sample.  Correcting for
the tagging and veto efficiencies obtained from MC with equal
fractions of $X(3940) \to D^{*0}\overline{D}{}^0$ and $X(3940) \to
D^{*+}{D}^-$ assumed, we find
\begin{eqnarray}
\mathcal{B}_{>2}(X(3940) \to D^{*}\overline{D}) =  \Xddst \nonumber \\
 (>45\%
\text{ at 90\% C.L.}),
\end{eqnarray}
where the systematic errors are taken into account for the lower
limit. In the limit of a vanishing fraction of low charged multiplicity
$X(3940)$ decays, the measured value of $\mathcal{B}_{>2}$ corresponds
to $\mathcal{B}(X(3940) \to D^{*}\overline{D})$.  

We set upper limits on the branching fractions of decay of $X(3940)$
into $D \overline{D}$ and $X(3940) \to J/\psi \omega$ final states,
taking into account the estimated systematic errors:
\begin{eqnarray}
\mathcal{B}(X(3940) \to D \overline{D}) < \Xdd  \text{ at 90\% C.L.;} \\
\mathcal{B}(X(3940) \to J/\psi \omega) < \Xom \text{ at 90\% C.L.}
\end{eqnarray}
These limits assume that low charged multiplicity $X(3940)$ decays are
negligible and, thus, may be overestimated.
  
In summary, we have observed a new charmonium-like state, $X(3940)$,
produced in the process $e^+e^-\rightarrow J/\psi \, X(3940)$, both in
inclusive production and via the $X(3940)\rightarrow D^*\bar{D}$ decay
mode. Both observations have a $5\sigma$ statistical significance.  We
have measured the Born cross section for the production process, the
branching fraction for $X(3940)\rightarrow D^*\bar{D}$, and set upper
limits on $X(3940)$ decays to $D\bar{D}$ and $J/\psi \omega$.

We thank the KEKB group for the excellent operation of the
accelerator, the KEK cryogenics group for the efficient operation of
the solenoid, and the KEK computer group and the NII for valuable
computing and Super-SINET network support.  We acknowledge support
from MEXT and JSPS (Japan); ARC and DEST (Australia); NSFC (contract
No.~10175071, China); DST (India); the BK21 program of MOEHRD and the
CHEP SRC program of KOSEF (Korea); KBN (contract No.~2P03B 01324,
Poland); MIST (Russia); MHEST (Slovenia); SNSF (Switzerland); NSC and
MOE (Taiwan); and DOE (USA).

\end{document}